\documentclass[12pt]{article}
\usepackage[utf8]{inputenc}
\usepackage[T1]{fontenc}
\usepackage{geometry}
\usepackage{amsmath,amssymb,amsfonts}
\usepackage[final]{microtype}
\usepackage{setspace}
\usepackage{caption}
\usepackage{graphicx}
\usepackage{hyperref}
\usepackage{enumitem}
\usepackage{tikz}
\usetikzlibrary{positioning,arrows.meta}
\title{Sybil-Resistant Service Discovery for Agent Economies}
\author{
  David~Shi \\ Operator~Labs \and
  Kevin~Joo \\ Operator~Labs
}
\date{Oct 31st, 2025}

\begin{document}
\maketitle

\begin{abstract}
x402 enables Hypertext Transfer Protocol (HTTP) services like application programming interfaces (APIs), data feeds, and inference providers to accept cryptocurrency payments for access. As agents increasingly consume these services, discovery becomes critical: which swap interface should an agent trust? Which data provider is the most reliable? We introduce TraceRank, a reputation-weighted ranking algorithm where payment transactions serve as endorsements. TraceRank seeds addresses with precomputed reputation metrics and propagates reputation through payment flows weighted by transaction value and temporal recency. Applied to x402's payment graph, this surfaces services preferred by high-reputation users rather than those with high transaction volume. Our system combines TraceRank with semantic search to respond to natural language queries with high quality results. We argue that reputation propagation resists Sybil attacks by making spam services with many low-reputation payers rank below legitimate services with few high-reputation payers. Ultimately, we aim to construct a search method for x402 enabled services that avoids infrastructure bias and has better performance than purely volume based or semantic methods. 
\end{abstract}

\section{Introduction}
Autonomous AI systems need machine-native payments to operate without humans in the loop. Legacy rails enforce subscriptions, delayed settlement, chargebacks, and manual invoicing, which block the real-time procurement of context, workflows, and compute. Browser usage reduce some friction but still assume human UX and credit cards. The x402 protocol~\cite{x402} closes the payments gap with onchain HTTP, so agents can pay per request and receive responses atomically (e.g., in USDC~\cite{usdc}). Our complementary goal is discovery: once agents can pay for any service, how do they decide \emph{which} to call?

Our central insight is to treat each payment as an endorsement whose strength depends on payer reputation, turning discovery into reputation-weighted ranking over the payment graph. We contribute: (i) a concise failure analysis of count/volume baselines and unseeded PageRank~\cite{brinpage98}; (ii) TraceRank, which seeds reputation and propagates it along value- and time-weighted flows (resolving the limits of manual TrustRank style ranking~\cite{gyongyi04}); and (iii) a minimal system that combines TraceRank with vector retrieval to return reliable services for agent queries.

\section{Background: x402 Payment-Gated Services}
x402 standardizes payment-gated HTTP endpoints: a client pays a service address, the service responds, and the payment is recorded onchain. Each payment forms a directed edge from payer to service with value and timestamp. This transparency enables discovery via revealed preference. 

Concretely, an initial request to a paid endpoint receives an HTTP 402 (Payment Required) describing terms; the client replays the request with a signed payment payload (e.g., an `X-PAYMENT` header). The server or an optional facilitator verifies the payment, triggers onchain settlement, and returns a 200 OK with the resource and an `X-PAYMENT-RESPONSE` containing settlement details. The protocol is stateless, HTTP-native, and chain-agnostic via facilitators, making per-request settlement practical for autonomous agents~\cite{x402}.

Perhaps the most compelling characteristic of the x402 protocol is the simplicity of integrating the client and server packages and its subsequent permissionless nature. Anyone can build a server, a facilitator, or a client, and with the increasing capability of large language models (LLMs), any agent can bootstrap the aforementioned with ease. 

\section{Why Existing Approaches Fail}
Counting payments invites Sybil spam and elevates infrastructure. Ranking by total volume rewards whales and enables wash trading. Both approaches over-index on quantity while ignoring who is paying and whether usage legitimate. 

Unseeded PageRank~\cite{brinpage98} on payment graphs promotes compulsory contracts (stablecoin issuers, automated market makers (AMMs), routers) rather than services, because transactions are protocol-mandated rather than editorial judgments. Manual seed lists (e.g., TrustRank~\cite{gyongyi04}) do not generalize in pseudonymous settings. None of these methods tie endorsements to who paid, how much, and when.

\subsection{The Spam Service Problem}
Consider two x402 services competing for discovery. \emph{Service A} (spam) advertises ``Send \$1, receive 1M airdrop'' and attracts 10{,}000 fresh wallets (airdrop farmers, bots), totaling 10{,}000 payments and \$10{,}000 volume. \emph{Service B} (legitimate) is a high-quality, specific background check service used by 50 sophisticated traders and protocols, totaling 50 payments and \$5{,}000 volume.

Under naive ranking, \emph{Service A} wins on transaction count (10{,}000 $\gg$ 50), on volume ($10$K $\gg$ $5$K), and on unseeded PageRank (more inbound edges). Popularity among low-reputation users outweighs endorsement by high-reputation users, inverting the discovery objective: agents searching for reliability see spam.

TraceRank weights each payment by payer reputation. If \emph{Service A}'s payers have near-zero seeds (fresh wallets), their collective endorsement contributes negligible reputation. If \emph{Service B}'s payers have high seeds (proven traders), each payment carries substantial weight. The outcome reverses: \emph{Service B} ranks higher despite lower raw counts and volume.

\section{TraceRank: Payments as Endorsements}
TraceRank operationalizes the payments-as-endorsements insight while resisting infrastructure bias and Sybil attacks. We precompute per-address reputation scores from external signals, then propagate those scores through economically and temporally weighted payment flows. Rather than counting transactions uniformly, each payment is weighted by the payer's seed reputation, value, and recency.

Let $\mathcal{V}$ denote addresses (users and services) and $\mathcal{E}$ denote directed payment edges. For each address $i\in\mathcal{V}$, assign a seed score $s_i$ from external data; addresses without seed data receive $s_i=0$. Over a chosen observation window, define aggregated flow
\begin{equation}
F_{j\to i} \,=\, \sum_{e\in\mathcal{E}_{j\to i}} \Big[ \log\big(1+\mathrm{value}_{\mathrm{USD}}(e)\big) \cdot e^{-\lambda\,\mathrm{age}_{\mathrm{days}}(e)} \Big],
\end{equation}
where $\log$ is natural and $\lambda$ has units of $\mathrm{day}^{-1}$ (ages measured in days). Let $S_i=\sum_k F_{k\to i}$ and define the normalized incoming-flow matrix $W$ by
\begin{equation}
w_{ji} \,=\, 
\begin{cases}
\dfrac{F_{j\to i}}{S_i}, & S_i>0, \\
0, & S_i=0~(\text{sinks}),
\end{cases}
\end{equation}
so that columns with positive inbound flow are exactly column-stochastic. TraceRank iterates
\begin{equation}\label{eq:tracerank}
\mathbf{r}^{(t+1)} \,=\, \mathbf{s} + \alpha\, W^{\top} \mathbf{r}^{(t)},\quad \alpha\in(0,1),
\end{equation}
which converges to the unique fixed point
\begin{equation}
\mathbf{r} \,=\, \big(I - \alpha W^{\top}\big)^{-1} \mathbf{s}.
\end{equation}
An immediate consequence is Sybil resistance: if a service receives payments from $N$ addresses with zero seed scores, it accumulates zero propagated reputation regardless of $N$. Bot wallets contribute no signal. Conversely, a single payment from a high-seed payer propagates meaningful reputation. This asymmetry makes fake popularity economically unappealing.
The score $r_i$ reflects both direct seed reputation and propagated endorsements along recent, value-weighted flows.

TraceRank is agnostic to seed provenance. Seeds may combine trading performance, decentralized social signals (e.g., Farcaster~\cite{farcaster}), protocol contributions, labeled entities (decentralized autonomous organizations (DAOs), verified protocols, funds), and agent attestations from ERC-8004 registries (identity, reputation, validation)~\cite{eip8004}. Log scaling, temporal decay, and normalization curb whales and wash trading, surface recency, and prevent infrastructure accumulation.

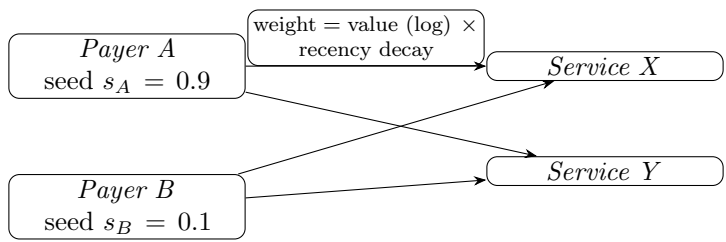
\begin{figure}[htbp]
\centering
\begin{tikzpicture}[
  node distance=10mm and 16mm,
  >=Stealth,
  every node/.style={draw, rounded corners, align=center, inner sep=2pt, font=\footnotesize, text width=3.0cm}
]
\node (pa) {\emph{Payer A}\\seed $s_A=0.9$};
\node (pb) [below=of pa] {\emph{Payer B}\\seed $s_B=0.1$};
\node (sx) [right=32mm of pa] {\emph{Service X}};
\node (sy) [below=of sx] {\emph{Service Y}};

\draw[->] (pa) -- node[above, sloped, font=\scriptsize]{$\text{weight} = \text{value (log)} \,\times\, \text{recency decay}$} (sx);
\draw[->] (pa) -- (sy);
\draw[->] (pb) -- (sx);
\draw[->] (pb) -- (sy);
\end{tikzpicture}
\caption{Reputation propagation: high-seed payers diffuse reputation along value- and time-weighted payment flows to services.}
\label{fig:propagation}
\end{figure}

\section{System Architecture for x402 Service Discovery}
We precompute a single TraceRank score per service from seeded accounts with value and time-weighted payment flows. Profiles are natural language descriptions of what the service does, and the vector embeddings of these profiles are precomputed. We use dense vector retrieval (optionally hybrid sparse + dense) to retrieve top-$K$ by cosine similarity and fuse with TraceRank via a simple multiplicative score:
\begin{equation}\label{eq:finalscore}
\mathrm{score}(A, q) \,=\, \cos(\mathbf{q}, \mathbf{p}_A) \times \mathrm{TraceRank}(A).
\end{equation}
This ranking returns the most semantically relevant services among those preferred by reputable payers. Agents can refine results by rephrasing or expanding the query and rerunning retrieval.

In an example PostgreSQL implementation with pgvector, store a per-service `tracerank` column and embeddings in one table, add vector and btree indexes, and compute the final score directly in SQL (e.g., `ORDER BY cosine(embedding, :q) * tracerank DESC`) with optional `WHERE` filters for chain, time window, or tags.

Counterfactual techniques include: (i) semantic-only, (ii) TraceRank-only, and (iii) volume and count oriented query techniques. Future work will demonstrate how the combined TraceRank and vector similarity technique excels at retrieving the highest quality services. 

\section{Conclusion}
Quality emerges from who pays, not just how much. TraceRank propagates precomputed reputation through value and time-weighted payment flows, producing a single score per service that resists Sybil spam, whale dominance, and infrastructure bias. 

Combined with semantic retrieval of service profiles, a simple multiplicative fusion yields a fast, agent-ready ranking. As agent economies scale, fast and high-quality service discovery becomes critical infrastructure. TraceRank shows that reputation can emerge from payment patterns and precomputed address related social reputation, enabling agents to bootstrap trust in decentralized marketplaces without privileged curators or identity systems.

\section*{Acknowledgements}
We thank SM Mesbahul Islam for his contributions to the reference implementation and valuable engineering support.

\end{document}